# Was the propeller mechanism active during the periastron passage of PSR 1259–63?

S. Campana[1,2], M. Colpi[3], S. Mereghetti[4], and L. Stella[1,2]

[1] Osservatorio Astronomico di Brera, Via Brera 28, I-20121 Milano, Italy;
e-mail: (campana, stella)@astmim.mi.astro.it
[2] Affiliated to I.C.R.A.
[3] Università degli Studi di Milano, Via Celoria 16, I-20133 Milano, Italy;
e-mail: colpi@astmiu.mi.astro.it
[4] Istituto di Fisica Cosmica del C.N.R., Via Bassini 15, I-20133 Milano, Italy;
e-mail: sandro@ifctr.mi.cnr.it

**Abstract.** Timing analisys of PSR 1259–63 during its recent periastron passage suggested that the pulsar was spinning-down due to the propeller mechanism (Manchester et al. 1995). This requires that the radio pulsar mechanism is temporarily quenched. On the basis of the Be equatorial disk model derived from the dispersion and rotation measures analysis (Melatos et al. 1995), we show that the mass inflow is sufficiently high to yield accretion down to the magnetospheric radius. In this case the X–ray spectrum might result from the propeller shock emission.

## 1. Introduction

PSR 1259–63 is the first discovered radio pulsar with a main sequence companion (Johnston et al. 1992). The modulation of the pulse arrival times ($P = 47.8$ ms) revealed a highly eccentric orbit ($e = 0.87$) with a period of $P_{orb} = 1237$ d (Johnston et al. 1994). The study of this system provides a unique opportunity to investigate the interplay between the radio pulsar and accretion phenomena, as well as to probe the structure of the Be star wind.

Melatos et al. (1995) modelled the time dependent radio dispersion and rotation measures across periastron, showing that the radio data could be explained equally well by a geometrically thin or thick stellar equatorial disk.

The pulse period measurements before and after the disappareance of the radio emission require the occurrence of a sudden spin-down episode at periastron. The most plausible interpretation of this period evolution is that the radio pulsar entered the propeller regime (Manchester et al. 1995; Ghosh 1995), i.e. the inflowing matter is stopped by the interaction with the rapidly rotating neutron star magnetosphere. This implies that the radio pulsar mechanism is turned off.

Here we show that only the thick Be equatorial disk can provide the mass inflow required to temporarily quench the radio emission. Furthermore, the resulting accretion rate is sufficiently small to avoid accretion down to the neutron star surface.

## 2. Radio pulsar activity versus accretion

Accretion onto the magnetospheric boundary takes place if the "pulsar radiation barrier" is overcome (see, for details, Campana et al. 1995). Equating the stellar wind ram pressure and the pulsar radiation pressure (which is a fraction $f$ of the spin-down luminosity) at the accretion radius, one finds for the case of PSR 1259–63

$$\dot M > \frac{f\,L^{sd}}{c\,v_{rel}} \simeq 2.8 \times 10^{18}\,f\,v_7^{-1}\ \mathrm{g\,s^{-1}}\,, \qquad (1)$$

where $\dot M$ is the mass capture rate and $v_7 = v_{rel}/10^7$ cm s$^{-1}$. Kulkarni & Hester (1988) showed that $f \sim 1$.

When the pulsar radiation barrier is won, the matter inflow can proceed inside the light cylinder radius ($r_{lc} = 2.3 \times 10^8$ cm), quenching the radio pulsar emission (e.g. Illarionov & Sunyaev 1975; Lipunov 1992). The motion of the infalling matter becomes dominated by the rapidly increasing magnetic field pressure at the magnetospheric boundary ($r_m \simeq 1.6 \times 10^8\,\dot M_{17}^{-2/7}$ cm, where $\dot M_{17}$ the accretion rate in units of $10^{17}$ g s$^{-1}$). This accretion down to the magnetospheric radius will produce, at least temporarily, a luminosity

$$L(r_m) = G\,M\,\dot M/r_m \simeq 8.0 \times 10^{36}\,f\,v_7^{-9/7}\ \mathrm{erg\,s^{-1}} \qquad (2)$$

(where $M$ is the neutron star mass), probably with a highly absorbed X-ray spectrum (in the spherical free-fall approximation $N_H \gtrsim 4 \times 10^{23}\,v_7^{-9/7}$ cm$^{-2}$). This implies also a very large free-free optical depth at radio wavelengths so that the transition to the magnetospheric accretion regime would occur when the radio pulsar is already undetectable.

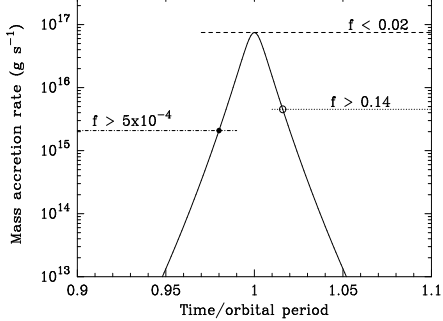

**Fig. 1.** Here we show the accretion rate experimented by the neutron star at periastron, adopting a thick disk model characterized by an exponential density profile both in the radial (scale radius $r_D \sim 12\,R_*$) and vertical directions (scale height $h_D \sim 0.6\,r_D$). The stellar matter density is $n_0 \sim 3 \times 10^8$ cm$^{-3}$ and its tempeature $\sim 10^4$ K. A disk tilting with respect to the orbital plane of $\theta \sim 10°$ has been adopted because it improves the fit of radio data (see for details Melatos et al. 1995). The filled circle signs the beginning of the radio signal eclipse and the open circle its end. The dashed line indicates the constraint on $f$ for the source to enter the propeller regime; the dot-dashed line the beginning of the radio eclipse and the dotted line the end of the eclipse.

The magnetospheric boundary expands for decreasing accretion rates. Eventually, $r_m$ becomes larger than $r_{lc}$ and the radio pulsar mechanism can resume. In the absence of accumulation of matter outside the magnetosphere, this is expected to take place for an accretion luminosity of

$$L(r_{lc}) \simeq 2.6 \times 10^{34} f \text{ erg s}^{-1} \ . \quad (3)$$

Due to its flatter radial dependence, the radio pulsar pressure sweeps the material outside the accretion radius and the pulsar pressure gives rise again to a shock front with the stellar wind. The expected column density is much smaller $N_H \sim 10^{22} f$ cm$^{-2}$. Note that the luminosity in Eq.(2) is substantially higher than that in Eq.(3), implying a higher luminosity threshold at the onset of accretion and, therefore, a limit cycle behaviour.

## 3. Propeller regime in PSR 1259−63

By modelling the time dependent radio dispersion and rotation measures, Melatos et al. (1995) derived the spatial distribution of the equatorial disk of the Be star. This allows us to compute the accretion rate experienced by PSR 1259−63 as a function of the orbital phase (Figure 1) for the two disks (geometrically thin or thick) compatible with the radio data.

In the case of an equatorial, geometrically thin Be disk, the radio pulsar would experiment an insufficient mass inflow rate to overcome the radiation pressure at periastron. The case of a thick equatorial disk seems more promising; however to establish if the $\dot M$ is sufficient to quench the radio pulsar emission requires some information on $f$ [cf. Eq.(1)]. This can be obtained by considering that (i) at the beginning of the radio eclipse the accretion rate must be lower than or equal to the threshold to win the pulsar pressure [cf. Eq.(1)] and that (ii) at the end of the eclipse the accretion rate must provide a luminosity lower or equal to that of Eq.(3). This yields $f \gtrsim 5 \times 10^{-4}$ and $f \gtrsim 0.1$, respectively.

For $f \gtrsim 0.1$, a geometrically thick disk gives a maximum accretion rate of $\sim 10^{17}$ g s$^{-1}$, high enough to win the pulsar pressure and enter the propeller regime.

## 4. Conclusions

We have shown that, the Be equatorial thick disk model compatible with the radio dispersion and rotation measures (Melatos et al. 1995), can provide a mass inflow on PSR 1259−63 near periastron sufficiently high to quench the radio emission and enter the propeller regime. It is interesting to note that, on the other hand, the mass inflow rate is small enough to avoid accretion onto the neutron star (which would produce an observable spin-up rather than the measured step-like spin-down).

In this scenario the X-ray emission seen with ASCA (Kaspi et al. 1995) and possibly OSSE (Grove et al. 1995) might result from accretion down to the magnetospheric radius (Stella et al. 1994; Campana et al. 1995), rather than from shock emission outside the accretion radius (Kaspi et al. 1995). The recent observation of pulsed X-ray emission from PSR 1259−63 near apoastron can also be better explained with emission from the neutron star rather than from a shock between the pulsar and Be star winds (Becker 1995). Although the observed X-ray luminosity below 10 keV can easily be provided by accretion down to $r_m$, the explanation of the power-law spectrum possibly extending in the hard X-ray range (Grove et al. 1995) would require the presence of non-thermal processes, resulting e.g. from a shock between the infalling plasma and the fastly rotating magnetosphere.